\documentclass[conference]{IEEEtran}
\renewcommand{\baselinestretch}{0.97}
\IEEEoverridecommandlockouts
\usepackage{cite}
\usepackage{tikz}
\usepackage{tcolorbox}
\usepackage{amsmath,amssymb,amsfonts}
\usepackage{algorithmic}
\usepackage{graphicx}
\usepackage{textcomp}
\usepackage{xcolor}
\usepackage{url}
\def\BibTeX{{\rm B\kern-.05em{\sc i\kern-.025em b}\kern-.08em
    T\kern-.1667em\lower.7ex\hbox{E}\kern-.125emX}}
\begin{document}

\title{Prompting GPT-5 on Scrum Certification Questions: An Empirical Accuracy Study
}


\author{
\IEEEauthorblockN{
Mirko Perkusich\textsuperscript{2},
Danyllo Albuquerque\textsuperscript{2}, 
João Paiva\textsuperscript{2},
Robson Vilar\textsuperscript{2},
Emanuel Dantas\textsuperscript{2}, \\
Ademar França de Sousa Neto\textsuperscript{2},
Rohit Gheyi\textsuperscript{1},
Kyller Gorg\^onio\textsuperscript{2},
and Angelo Perkusich\textsuperscript{2}
}
\IEEEauthorblockA{
\textsuperscript{1}\textit{Federal University of Campina Grande, Brazil}
}

\IEEEauthorblockA{
\textsuperscript{2}\textit{ISE Group, VIRTUS/UFCG, Campina Grande, Brazil}
}
\IEEEauthorblockA{
Corresponding author: emanuel.dantas@virtus.ufcg.edu.br
}
}

\maketitle
\begin{abstract}
Large Language Models (LLMs) are increasingly used in Agile Software Development for documentation, coaching, and training. As practitioners adopt these tools to prepare for certifications such as Professional Scrum Master (PSM), a key question is whether LLMs can reliably reason about Scrum, a framework with normative, well-defined rules described in the Scrum Guide (2020). This paper examines how different prompt techniques affect the factual accuracy of LLM responses to Scrum certification-style questions. A dataset of 993 validated PSM-aligned questions was answered by GPT-5 using three techniques: zero-shot, chain-of-thought, and with-source citation. All prompts achieved certification-level accuracy above 85\%, with the citation-based variant performing best (89.1\%) and yielding the lowest error rate. Correct answers concentrated in well-defined topics, such as \emph{Definition of Done}, Events, and Product Backlog Management, and in single-answer multiple-choice items, while multi-select questions and more interpretive areas, such as Scrum Team and Product Value, were less stable. Among questions where at least one prompt failed (16.2\%), errors clustered into misalignment with the Scrum Guide (28\%), content outside its scope (34\%), and outdated or biased interpretations (38\%). Overall, prompt techniques produced modest but consistent improvements, particularly in reducing misinterpretation and version drift, supporting more reliable use of LLMs in Agile learning and certification preparation.
\end{abstract}


\section{Introduction}
\label{sec:intro}

Agile Software Development has become the dominant approach to building and evolving software systems across industries \cite{patrucco2022scrum}. Its principles of collaboration, adaptability, and continuous improvement have shaped how teams design, build, and deliver software \cite{sutherland2018scrum}. As adoption has grown, so has the demand for formal education and certification programs that ensure a shared understanding of Agile values and frameworks \cite{crisostomo2024enhancing}. Certifications such as \textit{Professional Scrum Master (PSM)} and \textit{Professional Scrum Product Owner (PSPO)} from Scrum.org are widely used to assess practitioners' mastery of the Scrum framework and its underlying principles \cite{montenegro2019competences}.

Scrum, as codified in the \textit{Scrum Guide} (2020), defines normative and unambiguous rules governing accountabilities, events, and artifacts. This clarity makes Scrum-based certification exams a rigorous and measurable context for assessing knowledge consistency and factual understanding. Yet, learning and preparing for such certifications remains challenging, particularly for newcomers who lack access to experienced coaches or mentoring communities \cite{montenegro2019competences}.

Recently, Large Language Models (LLMs), such as ChatGPT, Gemini, and DeepSeek, have emerged as promising tools to support education and training \cite{Viegas_Gheyi_Ribeiro_2025,inoshita2024assessing}. They can generate explanations, quizzes, and feedback in natural language, potentially democratizing access to learning support. However, the reliability of their responses, especially when applied to normative, rule-based content such as Scrum, remains poorly understood. LLMs are known to produce plausible yet inaccurate statements, a phenomenon often referred to as ``hallucination,'' and their outputs are highly sensitive to the way prompts are formulated \cite{Shin2023PromptOrFineTune,SantanaJr2025WhichPromptingTechnique}. These limitations raise important concerns when such models are used in training or evaluative contexts.

In this paper, we investigate how different prompt techniques affect the correctness of GPT -5's answers to Scrum certification-style questions. We evaluate 993 English items aligned with the PSM I assessment format, which relies almost exclusively on textual prompts. Three prompt techniques are compared: (i) a \textit{zero-shot} baseline, (ii) a \textit{chain-of-thought} variant encouraging concise reasoning, and (iii) a \textit{with-source citation} prompt requiring explicit grounding in the Scrum Guide. All questions were executed independently through the OpenAI API to ensure reproducibility and prevent cross-contamination between responses.

By quantitatively comparing the accuracy of these prompt techniques, this paper advances understanding of how they affect the reliability of LLMs in structured, rule-based domains such as Scrum. Because our objective was to isolate the effect of prompt formulation without introducing variability from heterogeneous model behaviors, we deliberately relied on a single state-of-the-art model. The main contributions are threefold: first, we provide a curated and openly available dataset of 993 English, text-only Scrum certification-style questions, covering multiple knowledge areas and item formats aligned with PSM I certification; second, we introduce a reproducible empirical framework for evaluating LLM accuracy and consistency in normative Agile contexts, based on isolated question-answer executions and standardized scoring procedures; and third, we present a comparative assessment of prompt techniques, showing that structured and source-grounded prompts yield higher factual accuracy and reduce reasoning errors compared with the zero-shot baseline, while also deriving practical recommendations for Agile education and certification.

The remainder of this paper is organized as follows. Section~\ref{sec:background} reviews background concepts and related work on LLMs and AI-assisted Agile practices. Section~\ref{sec:design} describes the research design, dataset, and evaluation setup. Section~\ref{sec:results} presents and discusses the main findings. Section~\ref{sec:implications} outlines key implications. Finally, Section~\ref{sec:threats} discusses threats to validity, and Section~\ref{sec:final} concludes the paper with final remarks and future directions.

\section{Background and Related Work}
\label{sec:background}

This section reviews prior work on Scrum and its certification ecosystem, the use of LLMs in Agile and Scrum contexts, and the role of LLMs in educational and professional assessments. It concludes by identifying the gap that motivates our investigation into how accurately LLMs answer Scrum certification-style questions.

\textbf{Scrum and Agile Certification}. Scrum is one of the most widely adopted Agile frameworks worldwide, emphasizing iterative delivery, self-organizing teams, and continuous improvement \cite{sutherland2018scrum}. Its values, rules, and certification ecosystem are defined and maintained by Scrum.org, including the \emph{Scrum Guide} and professional certifications such as PSM and PSPO \cite{scrumorg2020}. Given its prescriptive and normative nature, Scrum provides a shared vocabulary and structured framework that helps organizations implement Agile practices consistently across teams and projects \cite{patrucco2022scrum}. These certification exams rely on multiple-choice questions derived from official Scrum materials, providing a controlled setting to evaluate knowledge accuracy. The PSM I, in particular, is an introductory-level certification composed of approximately 80 text-based questions to be completed in 60 minutes, with a minimum score of 85\% required to pass. These characteristics make Scrum certification tests an appropriate empirical context for examining how automated systems, such as LLMs, interpret and reproduce canonical Agile knowledge \cite{soroka2021importance}.

\textbf{LLMs in Agile and Scrum Contexts}. Recent studies have explored how LLMs can support agile software development by enhancing communication, documentation, and automation. Zhang et al.~\cite{zhang2024llm} applied an Autonomous LLM-based Agent System (ALAS) to improve user-story quality in industrial Agile teams, demonstrating gains in consistency and specification accuracy. Cabrero et al.~\cite{cabrero2024exploring} conducted an action-research study integrating AI assistants into Agile events, showing that such tools can strengthen collaboration and decision-making when organizational readiness is in place. Other studies investigated cognitive or multi-agent ecosystems that embed LLMs into Agile workflows. Chudziak et al.~\cite{chudziak2024towards} proposed the CogniSim framework, where LLM-powered agents assume roles such as developer or reviewer to coordinate Agile tasks. Cinkusz et al.~\cite{cinkusz2024cognitive} extended this concept to the Scaled Agile Framework (SAFe), reporting improvements in coordination and communication efficiency. More broadly, Modak et al.~\cite{modak2023integrating} showed that LLMs can accelerate coding, debugging, and documentation; Dhruva et al.~\cite{dhruva2024agile} proposed an LLM-based project management model to enhance visibility and delivery velocity; and Samimi et al.~\cite{samimi2025bridging} combined Model-Driven Engineering (MDE) and Scrum, using LLMs to automate modeling tasks and increase sprint adaptability. Collectively, these studies show growing interest in integrating LLMs into agile practices. However, most focus on productivity and collaboration, leaving open the question of whether LLMs can reliably reproduce the normative, rule-based knowledge codified in the \emph{Scrum Guide}.

\textbf{LLMs in Educational and Certification Assessments}. Beyond software development and Agile, recent research has examined how LLMs perform in formal educational and professional certification settings. Studies in medicine, academic admissions, and computing have shown that modern LLMs often achieve strong performance in well-structured, text-based exams. However, they still struggle with ambiguity, procedural reasoning, and visual prompts. For example, Gerard et al.~\cite{gerard2025evaluating} reported medical-exam performance at or above student level for leading models; Ashrafimoghari et al.~\cite{ashrafimoghari2024evaluating} found results above those of typical business-school candidates on the GMAT; Viegas et al.~\cite{Viegas_Gheyi_Ribeiro_2025} showed that several LLMs outperformed human candidates on text-based POSCOMP questions while still struggling with visual items; and Inoshita et al.~\cite{inoshita2024assessing} demonstrated that prompt techniques significantly affect accuracy in real-estate certification tasks. Together, these findings highlight both the potential and the limitations of LLMs as assessment and learning tools and reinforce the value of evaluating them in controlled, domain-specific certification contexts.

\textbf{Research Gap and Motivation}. Despite substantial progress in the field, little is known about how LLMs perform in Agile certification contexts, where correctness depends on strict adherence to the normative definitions of the \emph{Scrum Guide}. Prior LLM-Agile research has emphasized productivity and collaboration rather than evaluating whether models can reproduce canonical Scrum knowledge under exam-like conditions. Similarly, studies on LLMs in educational assessments seldom address domains with intentionally prescriptive rules and terminology. To bridge this gap, this study offers an empirical, prompt-based evaluation of LLM accuracy in a closed and highly structured Agile domain. By examining how different prompt techniques affect alignment with official Scrum principles, we provide new evidence on the reliability, limitations, and practical readiness of LLMs for supporting Agile education, training, and certification. Building on this gap, the following section describes the research design used to address it.

\section{Research Design}
\label{sec:design}

This section presents the study's methodological design, including the research goal, guiding questions, and the empirical process used to investigate how prompt techniques influence LLM accuracy on Scrum-style certification questions.

\subsection{Research Goal and Questions}
\label{subsec:goal-rq}

Guided by the Goal-Question-Metric (GQM) paradigm, our goal is: \textit{analyze GPT-5, in the context of Scrum fundamentals certification-style questions, for the purpose of evaluating how different prompt techniques affect its factual accuracy, topic-wise performance, and error behavior, from the perspective of Agile certification practitioners}. From this goal, we derive three research questions:

\begin{tcolorbox}[
  colback=black!2,
  colframe=black!35,
  title=Research Questions,
  coltitle=black,
  fonttitle=\bfseries,
  boxrule=0.4pt,
  arc=1.5pt,
  left=4pt,
  right=4pt,
  top=4pt,
  bottom=4pt
]

\textbf{RQ1.} What is GPT-5 accuracy under different prompt techniques?

\medskip

\textbf{RQ2.} What patterns of correct responses emerge across prompts, subject areas, and question formats?

\medskip

\textbf{RQ3.} What error patterns emerge across prompts, subject areas, and question formats?

\end{tcolorbox}

Together, these RQs examine overall factual correctness, variation across topics and question formats, and the main reasoning failures that remain despite prompt adjustments.

\subsection{Empirical Process}
\label{subsec:process}

\begin{figure}[t]
\centering
\scalebox{0.88}{
\begin{tikzpicture}[
  every node/.style={align=center, font=\scriptsize},
  process/.style={rectangle, rounded corners, draw=black, fill=gray!10,
                  text width=3.4cm, minimum height=0.95cm},
  arrow/.style={->, thick}
]

\node[process] (dataset) at (0,0) {1. Dataset Preparation\\(sources, schema, categories)};
\node[process] (prompting) at (4.2,0) {2. Prompt Execution\\(GPT-5 API, 3 prompts, isolation)};
\node[process] (evaluation) at (0,-1.8) {3. Answer Evaluation\\(validation, normalization, scoring)};
\node[process] (analysis) at (4.2,-1.8) {4. Comparative Analysis\\(accuracy, errors, patterns)};

\draw[arrow] (dataset) -- (prompting);
\draw[arrow] (dataset) -- (evaluation);
\draw[arrow] (prompting) -- (analysis);
\draw[arrow] (evaluation) -- (analysis);

\end{tikzpicture}
}
\caption{Overview of the empirical process.}
\label{fig:process}
\end{figure}
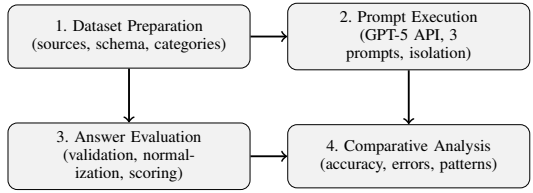

We follow a quantitative question-response-evaluation protocol inspired by prior LLM assessment work in software engineering and educational contexts \cite{Shin2023PromptOrFineTune,SantanaJr2025WhichPromptingTechnique,quille2024machine}. All experiments were conducted with GPT-5. Focusing on a single model allows us to isolate the effects of prompt formulation without introducing confounding variability from heterogeneous model behaviors. The study was organized into four stages, shown in Figure~\ref{fig:process}.

\textbf{Stage 1: Dataset Preparation}. The dataset originates from a professional Scrum certification course in which one author serves as an instructor. The questions are used to prepare candidates for the PSM I exam and have been answered and reviewed by thousands of learners worldwide, supporting both their coverage across Scrum subject areas and the correctness of items and answer keys. The final dataset comprises 993 questions.

Each item includes a question stem, alternatives, and a gold answer key. The dataset covers three common certification formats: \emph{True/False} (249 items), \emph{Multi-select} (153 items), and \emph{Multiple-choice} (591 items). Questions were stored in XLSX format using a minimal schema (\verb|question|, \verb|category|, \verb|gold_set|, \verb|options|). To ensure quality, two Scrum-certified reviewers cross-checked item correctness and category tags, and all texts were anonymized to remove course or vendor identifiers. The dataset and JSON schema are provided in the supplementary material.

\textbf{Stage 2: Prompt Execution}. Each question was evaluated independently, without conversational history or context carry-over, using the OpenAI API with the GPT-5 configuration. We compared three prompt techniques grounded in prior work on prompt engineering and empirical evaluation of LLMs \cite{Shin2023PromptOrFineTune,SantanaJr2025WhichPromptingTechnique,Viegas_Gheyi_Ribeiro_2025}: \textit{zero-shot}, which presents only the question and alternatives and serves as the baseline; \textit{chain-of-thought}, which requests a concise reasoning process before the final answer; and \textit{With-source citation}, which additionally requires explicit grounding in the \emph{Scrum Guide (2020)}. The interaction protocol was automated using a Python driver that loaded questions, executed all prompt variants, and stored responses in structured JSON/JSONL files, which were later consolidated into an XLSX file.

\textbf{Stage 3: Answer Evaluation and Data Handling}. Responses were validated and normalized before analysis. Records contained at least \verb|CorrectOption| or \verb|CorrectOptions|, and optionally \verb|brief_explanation| (or \verb|explication|) and \verb|reference|. We normalized answers by mapping lettered options to indices, extracting selected options from free text when necessary, and parsing multi-answer outputs into unordered sets. Accuracy was computed as the ratio between correctly selected options and the total number of correct options in the gold set, allowing proportional credit for partially correct multi-select answers. We also flagged empty outputs, off-schema JSON, and unsupported citations. Validated records were consolidated into an XLSX file for auditing and per-category analysis.

\textbf{Stage 4: Comparative Analysis}. We computed overall and by Scrum subject accuracy for each prompt technique, analyzed error frequencies and question-format patterns, and inspected option-selection distributions for positional bias. In addition, two authors collaboratively reviewed and classified erroneous cases to ensure a consistent interpretation of failure modes. Section~\ref{sec:results} reports these findings.

\section{Results and Discussion}
\label{sec:results}

This section presents the empirical results obtained from answering 993 Scrum certification-style questions under three prompt techniques: zero-shot, chain-of-thought, and with-source citation. Following the GQM plan defined in Section~\ref{sec:design}, we organize the results around the three RQs. Extended tables and raw analyses are available in the supplementary material.

\subsection{Overall Accuracy across Prompt Techniques (RQ1)}

\begin{figure*}[t]
\centering
\includegraphics[width=0.88\textwidth]{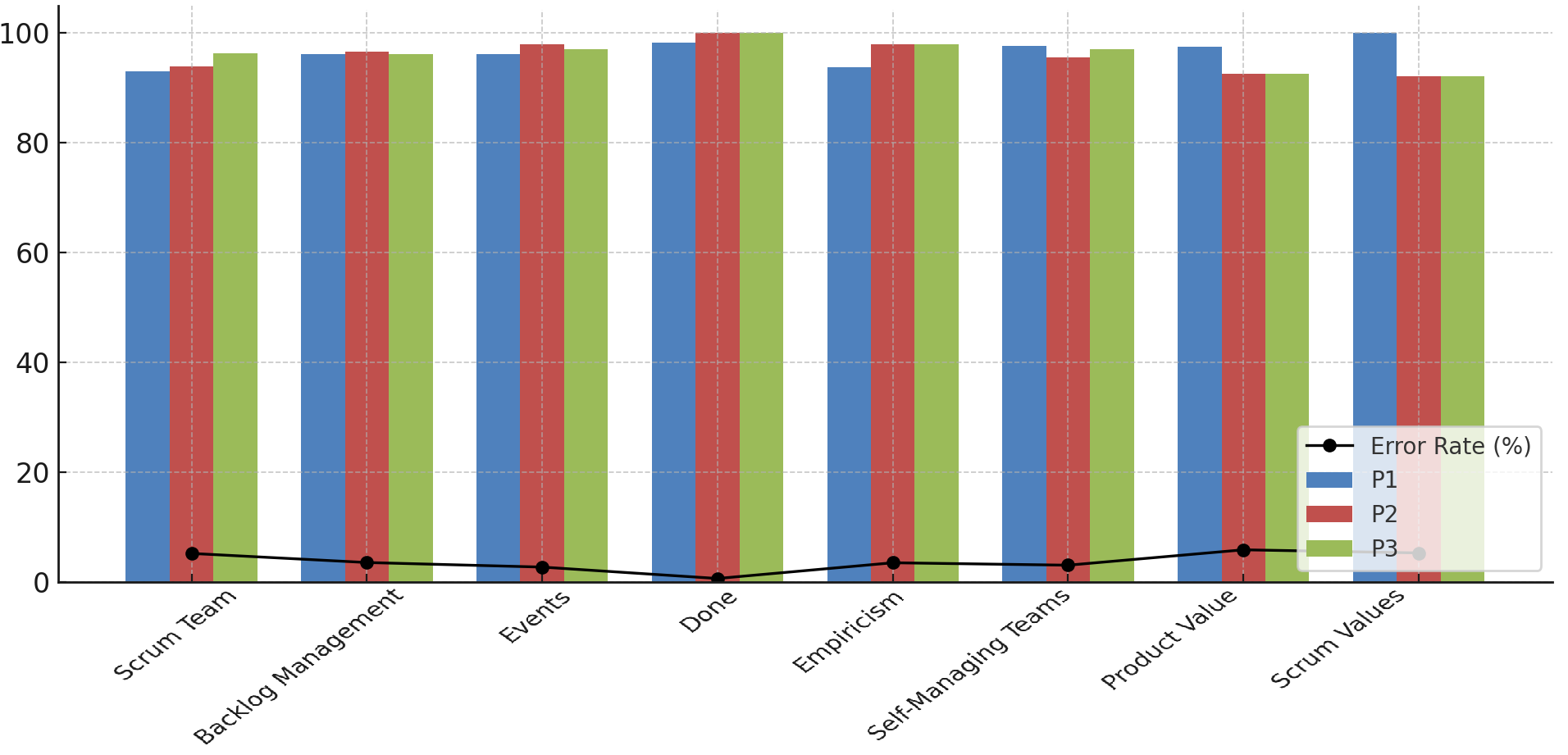}
\caption{Prompt accuracy and error rate by subject (Top 8). Bars represent mean accuracy per prompt (zero-shot [P1], chain-of-thought [P2], and with-source citation [P3]), while the black line denotes the overall error rate.}
\label{fig:prompt_accuracy_error}
\end{figure*}

The experiment was conducted in January 2026 using the model described in Section~\ref{subsec:process}. Fig.~\ref{fig:prompt_accuracy_error} summarizes mean accuracies for the eight most common subject areas, comparing prompt accuracy and error rates.

Across the full dataset, 920 of 993 questions (92.6\%) were answered correctly by at least one prompt technique, indicating consistently strong performance. Among the strategies, \emph{with-source citation} achieved the lowest error rates and the greatest stability across subjects and question types, whereas \emph{zero-shot} served as the most conservative baseline, with higher variability and a greater concentration of errors.

Foundational and well-defined Scrum areas, such as \emph{Definition of Done}, \emph{Events}, and \emph{Product Backlog Management}, approached perfect accuracy. In contrast, more interpretive domains, such as \emph{Scrum Team} and \emph{Product Value}, showed greater variation across prompts, consistent with their stronger reliance on contextual reasoning rather than definitional recall.

\begin{tcolorbox}[
  colback=white,
  colframe=black!30,
  title=Answer to RQ1,
  coltitle=black,
  fonttitle=\bfseries,
  boxrule=0.35pt,
  arc=1pt,
  left=4pt,
  right=4pt,
  top=4pt,
  bottom=4pt
]
All three prompt techniques achieved accuracy above 85\% on the PSM-I-aligned dataset. Among them, \emph{with-source citation} performed best (89.1\%), yielding the lowest error rate and the greatest stability across subjects and question types.
\end{tcolorbox}

\subsection{Patterns of Correct Answers (RQ2)}

To address RQ2, we analyzed the 920 questions for which at least one prompt produced a correct answer. Note that this set is not complementary to the 161 error cases examined in RQ3, since a question may be correctly answered by one prompt and incorrectly answered by another; consequently, both sets may overlap, and their combined count exceeds 993. Most successful predictions occurred in \emph{multiple-choice} items (565; 61.4\%), followed by \emph{True/False} (223; 24.2\%) and \emph{multi-select} questions (132; 14.3\%). This pattern suggests that the model performs best when selecting a single best option from a set of alternatives. At the same time, multi-select items remain more challenging due to the need for multi-label consistency.

Accuracy also varied slightly across prompt types. \emph{Zero-shot} achieved 876 correct predictions (87.9\%), \emph{chain-of-thought} achieved 880 (88.6\%), and \emph{with-source citation} achieved 885 (89.1\%). Only a very small fraction of responses were partially correct, indicating that outputs were usually either fully aligned or clearly incorrect. This progression suggests that minimal prompt engineering can produce small but meaningful gains in LLM-based assessments.

It is worth noting that \emph{Scrum Team} leads both in absolute hits (Table~\ref{tab:subject_hits}) and in absolute errors (Table~\ref{tab:error_subjects_compact}), which reflects its larger share of the dataset rather than contradictory performance; its high accuracy coexists with a high error count simply because it contains more questions than any other subject area.

Table~\ref{tab:subject_hits} summarizes the subjects contributing the largest number of correct answers.

\begin{table}[t]
\centering
\small
\caption{Subjects contributing the largest number of correct answers.}
\label{tab:subject_hits}
\begin{tabular}{lcc}
\hline
\textbf{Subject} & \textbf{Acc. (\%)} & \textbf{Hits} \\
\hline
Scrum Team                 & 94.4 & 232 \\
Product Backlog Management & 96.3 & 218 \\
Events                     & 97.1 & 149 \\
Definition of Done         & 99.4 & 55  \\
Empiricism                 & 96.5 & 48  \\
\hline
\end{tabular}
\end{table}

\begin{tcolorbox}[
  colback=white,
  colframe=black!30,
  title=Answer to RQ2,
  coltitle=black,
  fonttitle=\bfseries,
  boxrule=0.35pt,
  arc=1pt,
  left=4pt,
  right=4pt,
  top=4pt,
  bottom=4pt
]
Correct answers concentrated in multiple-choice questions and in well-defined normative topics, especially \emph{Definition of Done}, \emph{Events}, and \emph{Product Backlog Management}. 
\end{tcolorbox}

\subsection{Patterns of Errors and Failure Modes (RQ3)}

To address RQ3, we examined the 161 questions (16.2\% of the dataset) for which at least one prompt technique produced an incorrect answer. This subset is not complementary to the one used in the previous analysis, since a question may be correctly answered by one prompt and incorrectly answered by another. We first examine the distribution of errors across question formats, then analyze subject-level concentrations and classify the main failure modes. Within this subset, we observed 32 True/False and 43 multiple-choice questions. Although True/False items formed the smallest group, they were proportionally the most problematic, with low mean accuracies across prompts (21.9\% for zero-shot, 28.1\% for chain-of-thought, and 40.6\% for with-source citation). Their binary nature makes them highly sensitive to ambiguity, literal interpretation, and small semantic shifts. Multiple-choice questions showed moderate but stable performance, suggesting that prompts often recognized relevant concepts but struggled to discriminate among semantically close distractors. Multi-select questions achieved higher averages (59.8\% to 69.3\%) but also produced the largest number of partial errors, indicating incomplete relational reasoning.

Error counts also followed a clear gradient across prompts. \emph{Zero-shot} produced the largest number of full errors (90/161; 55.9\%), followed by \emph{chain-of-thought} (88/161; 54.7\%), while \emph{with-source citation} yielded the lowest count (80/161; 49.7\%). This pattern suggests that grounding answers in the Scrum Guide improves alignment with canonical definitions, although all prompts remain vulnerable to subtle linguistic and contextual shifts.

When disaggregated by subject, the largest concentrations of errors appeared in areas requiring procedural reasoning, role coordination, and context-dependent interpretation, as shown in Table~\ref{tab:error_subjects_compact}.

\begin{table}[t]
\centering
\small
\caption{Subjects contributing the largest number of errors.}
\label{tab:error_subjects_compact}
\begin{tabular}{lcc}
\hline
\textbf{Subject} & \textbf{Share of Errors (\%)} & \textbf{Errors} \\
\hline
Scrum Team                 & 28.0 & 45 \\
Product Backlog Management & 21.7 & 35 \\
Events                     & 13.0 & 21 \\
Definition of Done         & 6.2  & 10 \\
Self-Managing Teams        & 6.2  & 10 \\
\hline
\end{tabular}
\end{table}

\begin{tcolorbox}[
  colback=white,
  colframe=black!30,
  title=Answer to RQ3,
  coltitle=black,
  fonttitle=\bfseries,
  boxrule=0.35pt,
  arc=1pt,
  left=4pt,
  right=4pt,
  top=4pt,
  bottom=4pt
]
Errors were more frequent in questions requiring procedural reasoning, role coordination, and context-dependent interpretation, especially in \emph{Scrum Team}, \emph{Product Backlog Management}, and \emph{Events}. The main failure modes were factual misalignment with the Scrum Guide (28\%), content outside its explicit scope (34\%), and outdated or externally biased interpretations (38\%).
\end{tcolorbox}

\section{Implications}
\label{sec:implications}

This study has implications for Agile education, certification practices, and future research on LLM-based reasoning in structured knowledge domains.

For Agile education and training, the high overall accuracy observed, particularly in definitional and procedural topics, suggests that LLMs can be useful tools for supporting Scrum learning and exam preparation. However, the persistence of systematic errors in procedural and cross-role areas indicates that such tools should complement, rather than replace, critical human guidance. Their educational value depends on guided reflection, prompt literacy, and explicit grounding in normative references such as the \emph{Scrum Guide}.

For certification and professional practice, structured prompting proved important for maintaining alignment with canonical Scrum principles. Certification bodies and professional trainers may therefore use LLMs to generate questions, explanations, and formative feedback, provided that domain experts validate conceptual and procedural accuracy. Our results also show that even small prompt changes can affect reasoning fidelity, suggesting opportunities for adaptive certification systems and AI-assisted learning platforms. More broadly, the growing capabilities of LLMs in certification-style tasks reinforce the need for assessment providers to remain vigilant about exam integrity and appropriate safeguards in high-stakes settings.

From a research and benchmarking perspective, Scrum offers a controlled yet semantically rich setting for examining factual reliability and reasoning stability in LLMs. Future work should extend this benchmark to other Agile frameworks, such as Kanban, XP, and SAFe, to assess its generalizability across contexts that require different blends of declarative and situational reasoning. Combining quantitative metrics with expert-based qualitative assessment will also be important for understanding not only what LLMs answer correctly but also how and why reasoning errors occur. Overall, the results indicate that structured prompting can improve interpretability, accuracy, and pedagogical utility in normative Agile contexts. Moreover, these findings contribute to the software engineering community by demonstrating how LLM-based tools can be systematically evaluated and integrated into Agile development workflows, supporting more reliable human-AI collaboration in software teams.

\section{Threats to Validity}
\label{sec:threats}

As with any empirical study involving LLMs, this study faces threats that should be acknowledged.

Regarding construct validity, we operationalized ``accuracy'' as agreement between model outputs and predefined correct answers. Although appropriate for exam-style evaluation, this measure may underestimate reasoning quality in partially correct or conceptually valid responses, especially in multi-select and interpretive items. To reduce this risk, all questions and classifications were manually reviewed by multiple raters to confirm topical assignment and correctness. Even so, some forms of conceptual approximation and reasoning depth remain difficult to capture through automatic metrics.

Regarding internal validity, all prompts were executed independently, with conversation memory disabled to avoid cross-contamination between responses. Although stochastic factors such as temperature, token sampling, and backend variability may influence individual outputs, we mitigated these effects by using fixed inference parameters, consistent API settings, and full logging. A second threat concerns the correctness of the 993 gold-standard answers. While no dataset is entirely error-free, the items were sourced from a well-established PSM I preparation course used by thousands of learners and refined over multiple iterations, providing reasonable confidence in the answer key for comparative evaluation.

Regarding external validity, the dataset consists of Scrum certification-style questions explicitly aligned with the \emph{Scrum Guide} (2020). While this ensures normative fidelity, it limits generalization to other Agile frameworks, such as SAFe, LeSS, and Kanban, as well as to future revisions of the guide. Likewise, results are model- and configuration-specific; other architectures, fine-tuned variants, or newer models may yield different accuracy levels.

Regarding conclusion validity, our conclusions are based on a single domain-specific exam style and a single model, GPT-5, under fixed parameters. Prior research shows that LLM accuracy varies across exams and task types \cite{sallou2024breaking}, so the observed effects should not be generalized beyond Scrum without caution. Although the dataset is large and covers multiple Scrum topics, cross-framework and multi-model replications are still needed to confirm the robustness of the findings.

\section{Conclusion}
\label{sec:final}

This paper presents an empirical evaluation of how structured prompt techniques influence LLM accuracy on Scrum certification-style questions. Based on 993 validated items aligned with the \emph{Scrum Guide} (2020), we compared three prompt techniques to assess factual reliability and reasoning stability across question types and Scrum subject areas.

The results show that all three prompt techniques achieved accuracy above the 85\% PSM-I passing threshold (RQ1), confirming that GPT-5 would meet certification requirements regardless of prompt choice. Among them, the with-source citation prompt achieved the highest overall performance (89.1\%). Regarding patterns of correct answers (RQ2), accuracy was strongest in well-defined and normative areas, such as \emph{Definition of Done}, \emph{Events}, and \emph{Product Backlog Management}, as well as in single-answer multiple-choice questions. In contrast, multi-select items and more interpretive subjects, such as \emph{Scrum Team} and \emph{Product Value}, showed greater variability. The analysis of failure patterns (RQ3) revealed three dominant categories: factual misalignment with the Scrum Guide (28\%), content outside its explicit scope (34\%), and outdated or externally biased interpretations linked to pre-2020 practices (38\%). Together, these findings indicate that prompt structure has a modest but consistent impact on accuracy and error distribution, with source-grounded prompting reducing specific classes of mistakes rather than eliminating them.

Future work should extend this evaluation to other Agile frameworks, such as SAFe, LeSS, Nexus, and Kanban, and compare open and proprietary LLMs to assess how model differences affect normative alignment with the \emph{Scrum Guide}. Additional directions include metamorphic testing under controlled rewording, expert-based review, and longitudinal analyses across model versions and prompt techniques.

Taken together, these findings highlight both the potential and the limits of LLM-based reasoning in structured Agile domains. Although GPT-5 achieved certification-level accuracy, its recurring vulnerabilities in ambiguous situations show that Agile expertise cannot be reduced to textual reproduction. Effective use of LLMs in education, training, and certification, therefore, requires structured prompt design, human validation, and alignment with the normative intent of the Scrum Guide.

\section*{Artifacts Availability}
\label{sec-artifact}

Supplementary materials are available at: \url{https://github.com/Ilusinusmate/smxp26}.

\bibliographystyle{IEEEtran}
\bibliography{references}

\end{document}